\documentclass[manuscript,screen]{acmart}

\AtBeginDocument{%
  \providecommand\BibTeX{{%
    \normalfont B\kern-0.5em{\scshape i\kern-0.25em b}\kern-0.8em\TeX}}}

\setcopyright{none}
\copyrightyear{}
\acmYear{}
\acmDOI{}

\acmConference[]{}{}{}
\acmBooktitle{}
\acmPrice{}
\acmISBN{}

\begin{document}

\title{Kinetic Modeling of Magnetospheres}

\author{Stefano Markidis}
\affiliation{%
  \institution{KTH Royal Institute of Technology}
  \country{Sweden}
}

\author{Vyacheslav Olshevsky}
\affiliation{%
  \institution{KTH Royal Institute of Technology}
  \country{Sweden}
}

\author{Gabor T\'oth}
\affiliation{%
 \institution{University of Michigan}
 \country{MI, USA}
 }

\author{Yuxi Chen}
\affiliation{%
  \institution{University of Michigan}
  \country{MI, USA}
 }

\author{Ivy Peng}
\affiliation{%
  \institution{Lawrence Livermore National Laboratory}
  \country{CA, USA}
  }

\author{Giovanni Lapenta}
\affiliation{%
  \institution{KU Leuven}
  \country{Belgium}
  }
  
  \author{Tamas Gombosi}
\affiliation{%
  \institution{University of Michigan}
  \country{MI, USA}
  }

\renewcommand{\shortauthors}{Markidis, et al.}

\begin{abstract}
  This paper presents the state of the art of kinetic modeling techniques for simulating plasma kinetic dynamics in magnetospheres. We describe the critical numerical techniques for enabling large-scale kinetic simulations of magnetospheres: parameter scaling, implicit Particle-in-Cell schemes, and fluid-kinetic coupling. We show an application of these techniques to study particle acceleration and heating in asymmetric magnetic reconnection in the Ganymede magnetosphere.
\end{abstract}


\maketitle

\section{Introduction}
Kinetic simulations of plasma dynamics are the most powerful modeling tools for understanding the microscopic phenomena affecting global magnetospheric dynamics. Magnetic reconnection, collisionless shocks, kinetic instabilities, and particle-wave interaction are common magnetospheric phenomena occurring in very localized spatial regions. Nevertheless, they affect global magnetospheric dynamics. In particular, kinetic phenomena act as a localized seed for enabling large-scale events in the magnetosphere. Kinetic simulations can capture these localized phenomena and describe them correctly in a self-consistent way. Because of this, kinetic simulation tools are becoming an essential asset for any computational framework for physics-based space weather prediction~\citep{lapenta2013swiff,jordanova2018specification}. 

Kinetic simulations act as a microscope providing access to electron and ion distribution functions and phase-space densities. By analyzing this information, it is possible to understand particle acceleration and heating mechanisms. Humps in the distribution functions might reveal wave-particle interaction; holes in phase space might result from a microscopic instability. The most common simulation tools for studying magnetosphere dynamics are fluid or MHD simulations. They provide information about the evolution of distribution function moments together with the electromagnetic field. However, they fail in modeling the microscopic kinetic physics correctly. On the contrary, kinetic simulations address these fundamental limitations bridging global and kinetic scales in plasmas.

Kinetic modeling of magnetospheres comes at the cost of extreme computational demands. All the traditional kinetic simulation tools are affected by numerical stability constraints. These stability constraints require resolving electron spatial and temporal scales: grid spacing needs to be of the same order of Debye length and the time step needs to be a fraction of the plasma frequency. These scales are at least ten orders of magnitude smaller than global scales in the Earth magnetosphere. The problem's multi-scale nature makes the magnetosphere's global kinetic simulations extremely challenging, even when using the fastest supercomputers in the world. Three basic techniques make these simulations feasible today: plasma parameter scaling, implicit numerical schemes, and fluid-kinetic coupling. In this paper, we first describe these three techniques. We then show their application to study collisionless asymmetric magnetic reconnection in the Ganymede magnetosphere.

This article aims to review the current state of the art in fully kinetic simulations of the magnetospheres, presenting challenges and future directions. We focus on fully kinetic simulations where all the species, such as electrons, ions, and heavy ions, are modeled with kinetic description. For this reason, we do not cover hybrid models, where ions are modeled kinetically while electrons are modeled as fluid. We also do not describe how to model inner magnetosphere dynamics.

The paper is organized as follows. We present the governing kinetic equations in Section~\ref{eqs} and essential simulation techniques enabling the kinetic simulation of magnetospheres in Section~\ref{scaling1}. Section~\ref{app} presents an application of these techniques to study magnetic reconnection in the Ganymede magnetosphere. Finally, we discuss future directions in the field of kinetic simulations of magnetospheres in Section~\ref{conclusion}.

\section{Kinetic Equations for Magnetospheric Simulations} \label{eqs}
The kinetic behavior of a plasma species $\alpha$ (electrons or ions) is described by the distribution function $f_\alpha(\mathbf{x},\mathbf{v},t)$, providing the number density of plasma particles around the position $\mathbf{x}$ and velocity $\mathbf{v}$ in the six-dimensional phase space. The Vlasov equation, a conservation law for the phase-space density, governs the evolution of a collisionless plasma species $\alpha$ with mass $m_\alpha$ and charge $q_\alpha$ in the presence of electric and magnetic fields $\mathbf{E}$ and $\mathbf{B}$:
\begin{equation}
\label{Vlasov_kin}
\frac{\partial f_\alpha(\mathbf{x},\mathbf{v},t)}{\partial t} + \mathbf{v} \cdot \nabla_{\mathbf{x}} f_\alpha(\mathbf{x},\mathbf{v},t)+ \frac{q_\alpha}{m_\alpha} (\mathbf{E} + \mathbf{v} \times \mathbf{B}) \cdot \nabla_{\mathbf{v}} f_\alpha(\mathbf{x},\mathbf{v},t) = 0,
\end{equation}
where $t$ is time, $\mathbf{x}$ and $\mathbf{v}$ are the coordinates in the position and velocity spaces. Magnetospheric plasmas is fundamentally ``collisionless" because collisions occur at lower frequency than other dominant phenomena and they can be neglected. If the plasma to be modeled is collisional, an extra term is added to the right side of the Equation~\ref{Vlasov_kin}  and the equation would be then called ``transport" equation. These equations are called ``kinetic" because the distribution function depends on velocity. Resonance phenomena, such as wave damping mechanisms and kinetic instabilities, can be described correctly only if the velocity dependency is included in the model. 

On the other hand, fluid quantities, such as particle density ($n_{\alpha}$), fluid bulk velocity ($\mathbf{u}_\alpha$), and pressure ($p_\alpha$) describe the macroscopic plasma behavior. Basically, these quantities are weighted averages of the distribution function $f_\alpha(\mathbf{x},\mathbf{v},t)$ in the velocity space:
\begin{equation}
\label{moments}
\{ n_{\alpha}, n_{\alpha} {\bf u}_\alpha, p_\alpha \} = \int \{ 1, \mathbf{v}, m_\alpha (\mathbf{v} - {\bf u}_\alpha)(\mathbf{v} - {\bf u}_\alpha) \} f_\alpha(\mathbf{x},\mathbf{v},t) d{\bf v}.
\end{equation}

It is important to note that integral over the velocity eliminates the dependency on the velocity making fluid quantities dependent on the space variable. For this reason, fluid modeling cannot describe correctly all those phenomena whose accurate description requires to model the dependency on velocity. Ideal MHD models cannot account for collisionless dissipation mechanisms in magnetic reconnection and collisionless shocks. To solve Equation \ref{Vlasov_kin}, we need to calculate the electric and magnetic fields by solving Maxwell's equations. Equations \ref{Vlasov_kin}, \ref{moments} and Maxwell's equations constitute the Vlasov-Maxwell integral-differential system. The difficulty of solving such a system lies in the non-linear coupling between Vlasov and Maxwell's equations via integral equations expressing the distribution function moments. The solution of this integral-differential system in complex systems, such as a magnetosphere, is performed numerically.

\section{Enabling Technologies for Kinetic Simulations of Magnetospheres}\label{scaling1}

There are three broad numerical methods for solving the Vlasov-Maxwell system numerically: \textit{i}) Eulerian-Vlasov, \textit{ii}) spectral, and \textit{iii}) Particle-in-Cell (PIC) methods. The Euler-Vlasov method discretizes the phase space with an Eulerian grid. Spectral methods rely on the phase space's discretization using spectral functions such as Hermite and Legendre functions for the velocity and Fourier basis functions for space. Both Eulerian-Vlasov and spectral methods provide a highly accurate description of the evolution of distribution function. This feature is critical for modeling wave-particle interaction correctly. However, they require to discretize a six-dimensional phase space resulting in extremely high computational demands. Thus, these methods are not currently in use for the magnetosphere's fully kinetic simulations.

The third and most popular simulation technique for simulating the magnetosphere is the PIC method. The PIC algorithm's basic idea is to sample the phase space with a large number of computational particles and follow their trajectories in time. At each simulation time step, we reconstruct the distribution functions from the computational particles. Differently from na\"{\i}ve N-body simulations, PIC methods calculate the electric and magnetic field acting on the computational particles not directly by instead with the help of a grid. Charge and current densities are calculated on the grid using particle positions and velocities.  Maxwell's equations are solved numerically on the grid. Finally, the electric and magnetic fields on the grid are interpolated at particle position particles are advanced by using these electric and magnetic field values. The use of the grid in the PIC methods lowers the computational cost for calculating the fields from $\mathcal{O}(N_p^2)$ (na\"{\i}ve N-body simulations), where $N_p$ is the number of particles, to $\mathcal{O}(N_g \log N_g)$, where $N_g$ is the number of grid points. While PIC methods enable complex kinetic simulations at a relatively low computational cost, PIC simulations are affected by low accuracy that might impact wave-particle interaction modeling. The accuracy can be improved by increasing the number of particles (error decreases with $\sqrt{N_p}$) and increase the order of interpolation functions. In practice, most PIC simulations use linear interpolation functions and more than a hundred particles per cell to ensure the results' acceptable accuracy. Since the early Nineties, the PIC method is the de-facto standard tool for simulating magnetospheres with a kinetic approach. In particular, \citet{buneman1992solar} and \citet{nishikawa1997particle} have been then pioneers in exploring the feasibility of the first PIC global simulations of magnetospheres using the Tristan PIC code.

\subsection{Scaling of Plasma Parameters}
The kinetic simulation of planetary magnetospheres with realistic plasma parameters is currently an unfeasible task. In practice, numerical stability constraints and the large scale difference between electron and ion impose us to use a series of approximations for kinetic simulation of magnetospheres. Typically, we use a series of artificial scaling parameters in any current PIC simulation of the magnetosphere. Among the most important ones:
\begin{itemize}
\item \emph{Reduced Ion-to-Electron Mass Ratio.} It is a common practice in PIC simulations to reduce the ion to electron mass ratio to an artificial value. A reduced mass ratio allows compressing the temporal and spatial scales difference between electrons and ions. In fact, ion skin depth (or ion inertial length) and ion plasma period are $\sqrt{m_i/m_e} \times$ the electron skin depth and plasma period respectively. For this reason, setting $m_i/m_e$ to 100 allows us to gain a factor $4.3\times$ in grid spacing and time step and still retain a reasonable separation of scales. To lower the mass ratio further leads to even a larger gain. However, this comes with the risk of not separating enough electron and ion dynamics, impacting Hall physics modeling accuracy. Comparing magnetic reconnection simulations with realistic and reduced mass ratios shows that simulation results are qualitatively similar (same reconnection rate, outflow speed) in both cases. Simultaneously, electrostatic instabilities are more present in simulations with a realistic mass ratio~\citep{lapenta2011bipolar}. We note that most PIC simulations use a reduced ion to electron mass ratio outside the field of magnetospheric kinetic simulations.
\item \emph{Increased Solar Wind Velocity.} Kinetic simulations of magnetosphere require to cover a time so that solar wind particles initially injected from one boundary can reach the other boundary on the opposite side. This time is called ``transition time," and it is necessary to observe the full formation of a magnetosphere in the simulation. Thus, a simulation of the magnetosphere needs to cover at least one transition time. To use artificially high solar wind velocity leads to shorter transition times and consequently to a smaller number of simulation cycles. The first kinetic simulations of magnetosphere by~\citet{buneman1992solar} and~\citet{nishikawa1997particle} used a solar wind velocity value that is half the speed of light in the vacuum. This value is roughly $750\times$ the realistic solar wind speed. Because of such a high value, we must use relativistic equations in these simulations. More recent kinetic simulations of magnetospheres \citep{peng2015formation} use solar wind velocity that is 2\% the speed of light, still $15\times$ larger than the realistic value.
\item \emph{Kinetic Scaling Factor.} 
An additional ``kinetic scaling" factor $f=d_g/d_i$, where $d_g$ is the magnetopause stand-off distance and $d_i$ is the ion skin depth, has been proposed to model kinetically large magnetospheres~\citep{toth2017scaling}. The kinetic scaling is used when the global scales (magnetopause standoff distance) are much larger than the kinetic scales (ion skin depth). When using this scaling, we increase the ion and electron mass to charge ratios by $f$. The computational cost is reduced as the scaling factor's third and fourth powers in two- and three-dimensional simulations. The kinetic scaling is crucial for simulation of large magnetospheres, such as Earth,~\citep{chen2017global}, Saturn and Jupiter magnetospheres.
\end{itemize}

\subsection{Implicit Particle-in-Cell Schemes} 
When it comes to simulating magnetospheres, it is convenient to use the largest time step and grid spacing, still retaining numerical stability and an accurate micro-physics description. Conventional explicit PIC methods have severe stability constraints. On the other hand, an implicit time-discretization of the PIC equations allows for relaxing the numerical stability constraints at increased computation costs. Implicit PIC algorithms enable simulations with time steps that are the order of the plasma period and grid spacing that are 10-100 Debye lengths~\citep{markidis2010multi}. The use of implicit PIC methods impacts the dispersion relation of waves in the system. In particular, implicit PIC algorithms introduce \emph{spectral compression} and \emph{selective damping}. If waves have a frequency higher than the simulation Nyquist frequency, their frequency is artificially lowered to the Nyquist frequency. This implicit PIC method mechanism is called ``spectral compression". If the simulation time step does not resolve the wave period, the wave is damped (``selective damping"). Implicit PIC codes enable 3D kinetic simulations of magnetospheres with simulation boxes that are hundreds ion skin depth in size in each direction \citep{peng2015formation,peng2015kinetic}. However, despite all the algorithmic improvements of implicit PIC methods, global PIC simulations can still only simulate a relatively small region of large magnetospheres.

\subsection{Coupling Fluid and Kinetic Algorithms}
Plasma behaves like a fluid in large regions of magnetospheres. Modeling regions where the fluid approximation holds do not require computationally expensive kinetic simulations. Instead, a kinetic description is needed for simulating regions where magnetic reconnection and collisionless shocks occur. For this reason, the most recent numerical methods for plasma simulation combine kinetic and fluid descriptions within the same framework. This technique is called ``fluid-kinetic coupling". The most successful formulation of fluid-kinetic coupling is the MHD-EPIC (MHD with Embedded PIC) method~\citep{daldorff2014two}. This method allows us to have one or more PIC localized regions overlapping with a much larger MHD simulation box. The MHD-EPIC fluid-kinetic coupling is \emph{two-way}: the MHD simulation provides the PIC regions with boundary conditions. The small PIC regions provide the MHD simulation with the whole domain grid values. 
Typically, the MHD simulation is run first without embedded PIC regions. This MHD simulation is to form the whole magnetosphere structure. Then, we introduce one or more PIC regions in localized regions where phenomena such as magnetic reconnection and collisionless shock are expected to occur. In the PIC regions, the electromagnetic field is initialized using values from the MHD region, and particles are initialized with a Maxwellian distribution. We calculate the Maxwellian distribution's drift and thermal velocities from the MHD simulation current and pressure values. 

\begin{figure}[ht]
\begin{center}
\includegraphics[width=\columnwidth]{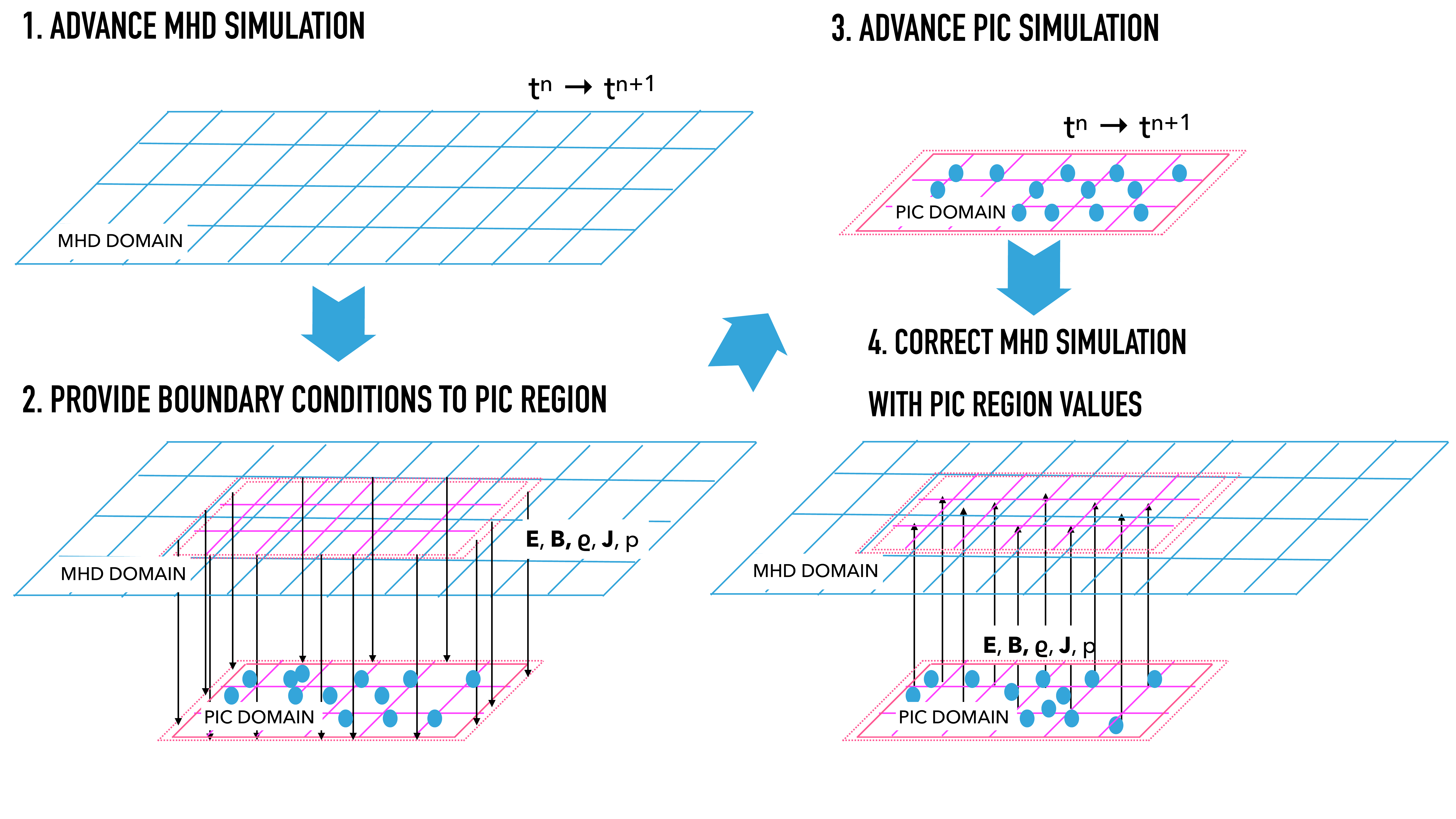}
\caption{The four steps of the MHD-EPIC computational cycle, coupling fluid and kinetic simulations.}
\label{setupSim}
\end{center}
\end{figure}

After the PIC region initialization, the MHD-EPIC method repeats four basic steps at each computational cycle. We summarize these four steps in Figure~\ref{setupSim}. First, an MHD step is performed all over the domain. Second, the MHD quantities are imposed as boundaries of the PIC regions. Third, we carry out a PIC update. Finally, the MHD quantities are completely replaced by the PIC simulation quantities.

While multiple PIC steps can run for one MHD step as a form of \emph{kinetic sub-cycling}, the MHD and kinetic steps are synchronized at each time step to avoid inconsistency between fluid and kinetic moments~\citep{markidis2014fluid}. Since MHD and PIC simulations use the same time step, it is crucial to employ an embedded implicit PIC method to use large time steps.

\section{Example Use Case: Asymmetric Magnetic Reconnection in Ganymede Magnetosphere}\label{app}

This section presents a kinetic magnetosphere simulation combining all the critical technologies described previously. We focus on simulating collisionless asymmetric reconnection in Ganymede magnetosphere as in Ref.~\citep{toth2016extended}. Differently, we use only one PIC region that is localized in the region of asymmetric reconnection and has a higher spatial resolution. Also, we focus on studying particle distribution functions and phase space.

\subsection{Computational Framework} 
The MHD-EPIC algorithm and code is part of the Space Weather Modeling Framework (SWMF). In essence, the MHD-EPIC implementation is the software coupling of two principal parallel codes for space physics simulations: the BATS-R-US~\citep{powell1999solution} and iPIC3D~\citep{markidis2010multi, peng2015energetic} codes. The two codes have been adapted as computer libraries exposing their main computational steps as library functions. An additional library, called SWMF, takes care of the data movement and communication between BATS-R-US and iPIC3D codes. Therefore, the MHD-EPIC implementation is a stand-alone code using the BATS-R-US, iPIC3D, and SWMF libraries. 

In specific, BATS-R-US is a Fortran90/MPI finite volume library for solving MHD and extended MHD with block-adaptive technique. iPIC3D code is a highly optimized C++/MPI implicit PIC library for 3D Cartesian geometries. The SWMF is a Fortran90/MPI library handling the two libraries' process placement and the data movement between BATS-R-US and iPIC3D. 

The MHD-EPIC code requires the usage of supercomputers. While the computational cost of MHD-EPIC simulations largely varies depending on the type and size of the problem and the number of PIC regions, we note that the MHD-EPIC simulation we present in this paper runs for one day 2,048 cores on a Cray XC40 supercomputer. We also note that the PIC simulation took 88\% of the total computational cost despite using only one PIC region in a tiny part of the domain.

\subsection{Simulation Setup} 
We perform 3D MHD-EPIC simulations of the Ganymede magnetosphere with a PIC region covering the region where asymmetric reconnection occurs. In the simulation reference system, the Jovian wind flows in the simulation box in the positive X direction. The XY plane is the equatorial plane, and the XZ plane is the meridional plane. 

As an initial set-up, we mimic Ganymede's magnetosphere's plasma parameters as observed by Galileo G8 flyby in 1997. Ganymede is exposed to the Jovian wind that has a particle density of $n = 4/cm^3$, a velocity $v = 140 km/s$ and pressure $p =3.8 nPa$. In our simulation, the magnetic field convected by Jovian wind points Southward, $\mathbf{B} = (0, -6, -77) nT$. Ganymede dipole magnetic field is $750 nT$ at the equator on the surface. 

In this configuration, the Jovian wind is both sub-sonic and sub-Alfv\'enic. For this reason, the interaction between Jovian wind and Ganymede magnetic dipole results in the characteristic Alfv\'enic wings. The BATS-R-US MHD simulation uses an adaptive Cartesian grid with the smallest grid cell equal to $1/32 R_G$ where $R_G$ is the radius of Ganymede, with a total of 8.5 million cells. The MHD simulation adaptive grid discretizes a cube box with $256 R_G$. In the PIC region's proximity, the Hall term in extended MHD discretized equations is on to allow the fluid-kinetic coupling~\citep{daldorff2014two}. 

A single PIC region with size $L_x= 0.5 R_G$, $L_y=1.5 R_G$ and $L_z=1.25 R_G$ is introduced. The PIC grid consists of $96 \times 288 \times 244$ points with a grid spacing of $1/192 R_G$. The size of the PIC simulation box is determined by taking into account two factors. We first need to ensure that grid spacing is small enough to capture the electron dynamics with acceptable accuracy. Second, the choice of the total number of cells is limited by the computational resources in use. We do not use kinetic scaling as there is no large separation of global and kinetic scales in the Ganymede magnetosphere ($f \approx 10$). However, we scale the ion to electron mass ratio to an artificial value of 100. We use 432 particles per cell. 

\begin{figure}[ht]
\begin{center}
\includegraphics[width=1\columnwidth]{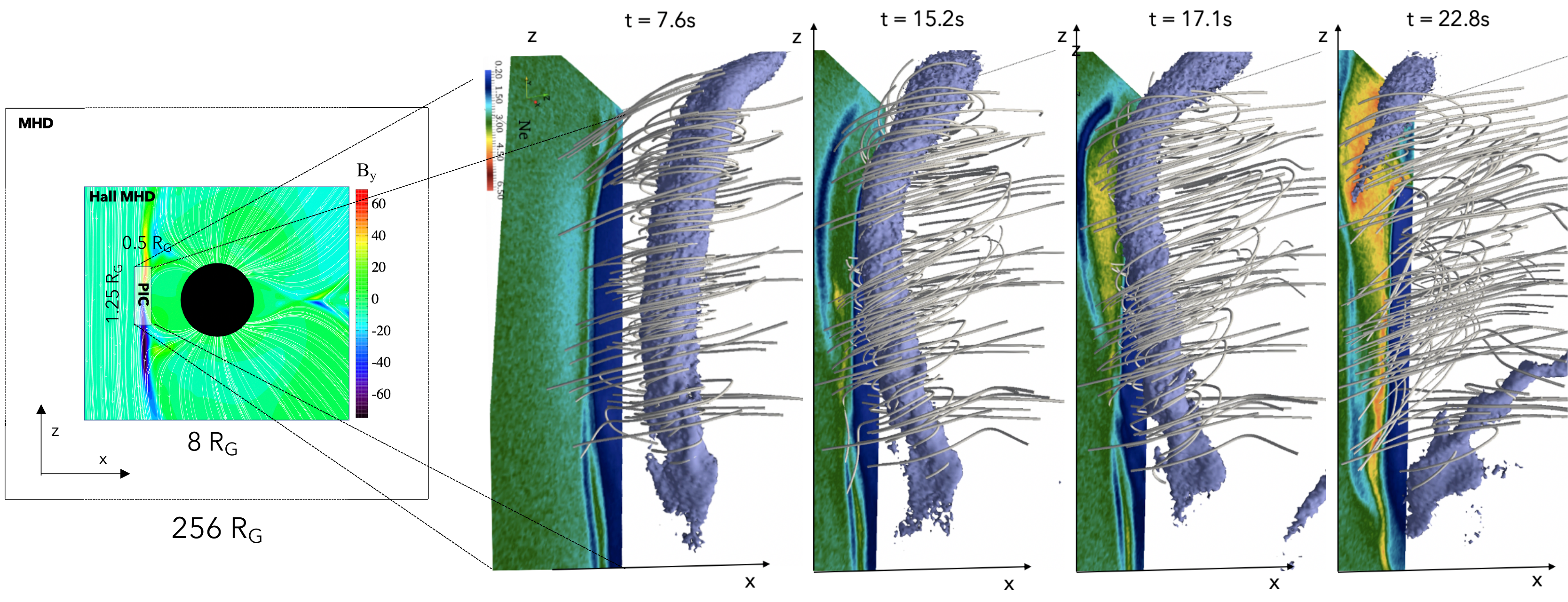}
\caption{The PIC region is embedded in larger Hall MHD and MHD boxes (left panel). The right panel shows the evolution of a flux ropes (high density iso-surfaces) and magnetic field lines (grey tubes) at different times.}
\label{setup}
\end{center}
\end{figure}

The average time step is 0.001 seconds (same value for MHD), and we run the simulation over 18,000 computational cycles covering a physical time of 25 seconds. Figure~\ref{setup} shows the different simulation MHD, Hall MHD, and PIC boxes on the left panel. We show the simulation evolution in the PIC region on the right panels: high electron density surfaces are represented as iso-surfaces in light blue color while the grey lines represent the magnetic field lines. A contour plot at the bottom of the PIC region represents the electron density.  The plots show the dynamics of a flux rope during collisionless asymmetric reconnection.

\subsection{Electron and Ion Distribution Functions} 

One of the advantages of kinetic simulations is the possibility of accessing electron and ion microscopic information. While Figure~\ref{setup} presents quantities that are in principle also available from MHD codes, self-consistent distribution functions are only available from kinetic simulations. 

For instance, we analyze the distribution functions in the proximity of the asymmetric magnetic reconnection X point. We first identify the reconnection point on one of the XZ planes by plotting the parallel electric field. In the simulation, the electron diffusion region is approximately 300 km (approximately eight electron skin depth) long. 

\begin{figure}[ht]
\begin{center}
\includegraphics[width=1\columnwidth]{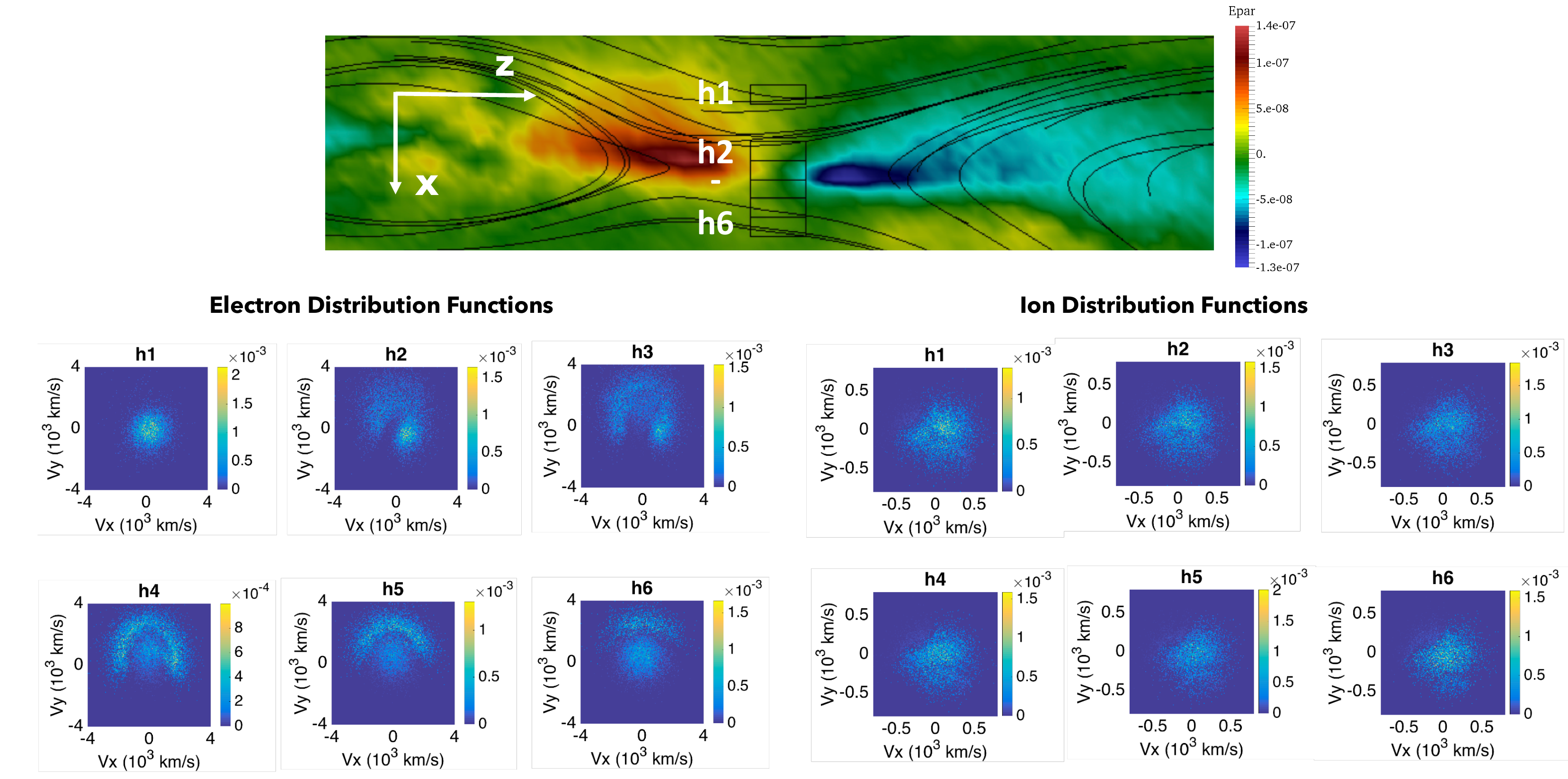}
\caption{Electron and ion distribution functions in different spatial regions h1-h6 close to X-line during asymmetric magnetic reconnection in Ganymede magnetosphere. An analysis of the electron distribution functions shows the crescent distribution function in proximity of the X point (boxes h4-h6). The top panel represents a contour plot of the parallel electric field with superimposed magnetic field lines in proximity of the X point.}
\label{diffusionregionPS}
\end{center}
\end{figure}

To calculate the distribution functions, we count all the electrons or ions in small regions of space (indicated as h1-h6 in Figure~\ref{diffusionregionPS}) and perform a two dimensional binning in the Vx and Vy directions. The number of particles per bin is then divided by the region's total number and plotted as a contour plot. When investigating the electron distribution functions in Figure~\ref{diffusionregionPS}, we find crescent distribution functions in the proximity of the X point (boxes h4-h6).

\subsection{Phase-Space Density} 
\begin{figure}[ht]
\begin{center}
\includegraphics[width=1\columnwidth]{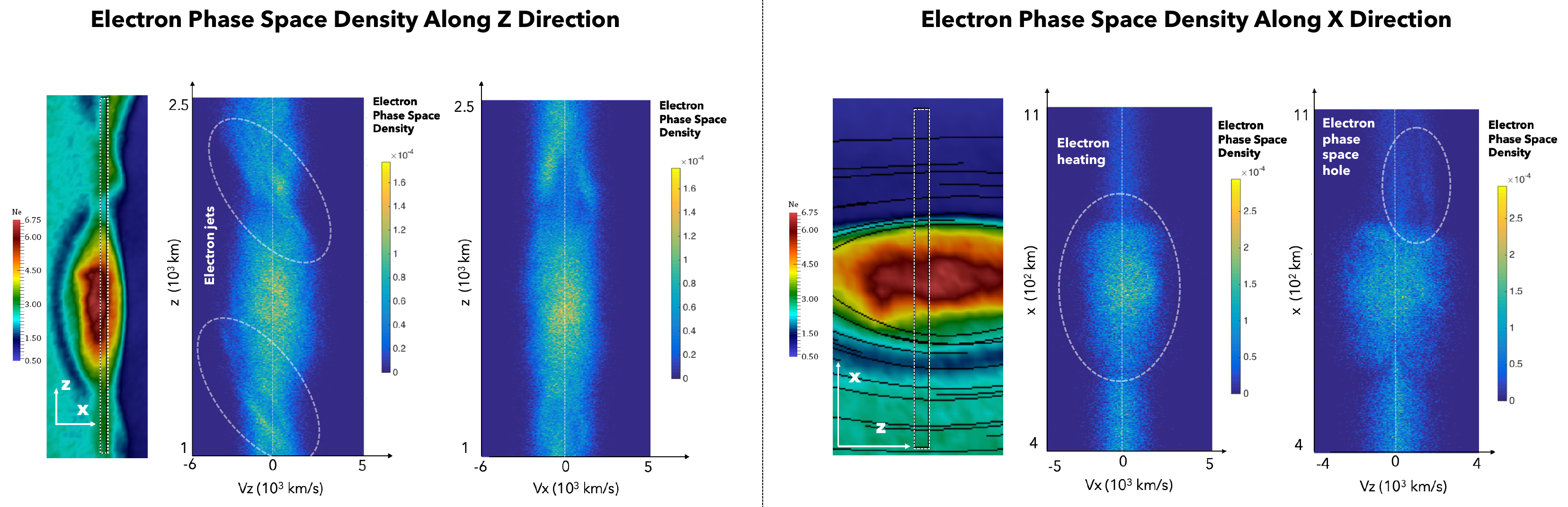}
\caption{Electron phase space density along the Z and X directions. An analysis of electron phase space density shows electron jets in proximity of the X points, electron heating within the flux rope and an electron phase space hole in proximity of magnetic reconnection separatrices.}
\label{phasespace}
\end{center}
\end{figure}

Another important quantity that is available from kinetic simulations is the electron and ion phase-space density. The analysis of these quantities allows us to understand acceleration and heating mechanisms and identify microscopic instabilities. We reconstruct the electron phase space density in a similar way we reconstructed the distribution functions. We show it in Figure \ref{phasespace}. In the left panel, the phase space is reconstructed along the Z direction crossing two X points in the flux rope. 

This electron phase space density shows two pairs of electron jets exiting the two X points (enclosed in the two dashed ellipses). The right panel of Figure~\ref{phasespace} shows the electron phase space density along the X direction entering and exiting the flux rope. Electrons are heated within the flux rope. Also, an electron phase space hole is visible in the electron phase space density, hinting at an instability that could have generated the electron space hole along the magnetic reconnection separatrices.

\section{The Future of Kinetic Simulations of Magnetospheres}\label{conclusion}

This article described the current fundamental numerical techniques enabling large-scale kinetic simulations of magnetospheres. Kinetic simulations allow us to access electron and ion micro-physics and identify particle acceleration and heating mechanisms.

The MHD-EPIC is currently the most powerful tool for enabling the kinetic simulation of magnetospheres. It has been extremely successful in modeling planetary magnetospheres from Ganymede \citep{toth2016extended}~\citep{newGanymede} to Mars~\citep{ma2018reconnection}, Mercury~\citep{newMercury} and Earth~\citep{chen2017global}. However still, some challenges remain. One challenge is to initialize particle distribution functions using a kinetic equilibrium configuration in the PIC regions. In fact, the distribution functions satisfying a kinetic equilibrium in our simulations' electromagnetic configuration is not known. In our simulations, PIC particles are initialized with a Maxwellian distribution function that might not be the kinetic equilibrium distribution function. This approximation might introduce artificial effects in the initial part of the simulation. The impact of this approximation on the simulation results is not clear, and further studies are required. A similar challenge occurs at the PIC regions boundary where particles are injected assuming a Maxwellian distribution. To use Maxwellian distribution functions is an acceptable approximation if the PIC region boundary is far from the space where the phenomena of interest occur. However, they might not be appropriate if energetic events leading to non-Maxwellian distributions occur close to the boundary. The last limitation of current MHD-EPIC implementations is the fact that PIC regions are static.  It is important to move the PIC regions during the simulation. An alternative approach would be adopting a numerical scheme that allows for smoothly transitioning from fluid to kinetic description only when required. Novel numerical methods, such as the PolyPIC~\citep{polypic} and Hermite-Fourier spectral methods~\citep{vencels2015spectral} have a built-in capability for solving this problem, but they are still in their initial development.

\bibliographystyle{ACM-Reference-Format}
\bibliography{sample-manuscript}


\begin{thebibliography}{20}


\ifx \showCODEN    \undefined \def \showCODEN     #1{\unskip}     \fi
\ifx \showDOI      \undefined \def \showDOI       #1{#1}\fi
\ifx \showISBNx    \undefined \def \showISBNx     #1{\unskip}     \fi
\ifx \showISBNxiii \undefined \def \showISBNxiii  #1{\unskip}     \fi
\ifx \showISSN     \undefined \def \showISSN      #1{\unskip}     \fi
\ifx \showLCCN     \undefined \def \showLCCN      #1{\unskip}     \fi
\ifx \shownote     \undefined \def \shownote      #1{#1}          \fi
\ifx \showarticletitle \undefined \def \showarticletitle #1{#1}   \fi
\ifx \showURL      \undefined \def \showURL       {\relax}        \fi
\providecommand\bibfield[2]{#2}
\providecommand\bibinfo[2]{#2}
\providecommand\natexlab[1]{#1}
\providecommand\showeprint[2][]{arXiv:#2}

\bibitem[\protect\citeauthoryear{Buneman, Neubert, and Nishikawa}{Buneman
  et~al\mbox{.}}{1992}]%
        {buneman1992solar}
\bibfield{author}{\bibinfo{person}{Oscar Buneman}, \bibinfo{person}{Torsten
  Neubert}, {and} \bibinfo{person}{K-I Nishikawa}.}
  \bibinfo{year}{1992}\natexlab{}.
\newblock \showarticletitle{Solar wind-magnetosphere interaction as simulated
  by a 3-D EM particle code}.
\newblock \bibinfo{journal}{\emph{IEEE transactions on plasma science}}
  \bibinfo{volume}{20}, \bibinfo{number}{6} (\bibinfo{year}{1992}),
  \bibinfo{pages}{810--816}.
\newblock


\bibitem[\protect\citeauthoryear{Chen, T{\'o}th, Cassak, Jia, Gombosi, Slavin,
  Markidis, Peng, Jordanova, and Henderson}{Chen et~al\mbox{.}}{2017}]%
        {chen2017global}
\bibfield{author}{\bibinfo{person}{Yuxi Chen}, \bibinfo{person}{G{\'a}bor
  T{\'o}th}, \bibinfo{person}{Paul Cassak}, \bibinfo{person}{Xianzhe Jia},
  \bibinfo{person}{Tamas~I Gombosi}, \bibinfo{person}{James~A Slavin},
  \bibinfo{person}{Stefano Markidis}, \bibinfo{person}{Ivy~Bo Peng},
  \bibinfo{person}{Vania~K Jordanova}, {and} \bibinfo{person}{Michael~G
  Henderson}.} \bibinfo{year}{2017}\natexlab{}.
\newblock \showarticletitle{Global three-dimensional simulation of {Earth's}
  dayside reconnection using a two-way coupled magnetohydrodynamics with
  embedded particle-in-cell model: Initial results}.
\newblock \bibinfo{journal}{\emph{Journal of Geophysical Research: Space
  Physics}} \bibinfo{volume}{122}, \bibinfo{number}{10} (\bibinfo{year}{2017}),
  \bibinfo{pages}{10--318}.
\newblock


\bibitem[\protect\citeauthoryear{Chen, Toth, Jia, Slavin, Sun, Markidis,
  Gombosi, and Raines}{Chen et~al\mbox{.}}{2019}]%
        {newMercury}
\bibfield{author}{\bibinfo{person}{Yuxi Chen}, \bibinfo{person}{Gabor Toth},
  \bibinfo{person}{Xianzhe Jia}, \bibinfo{person}{James Slavin},
  \bibinfo{person}{Weijie Sun}, \bibinfo{person}{Stefano Markidis},
  \bibinfo{person}{Tamas Gombosi}, {and} \bibinfo{person}{Jim Raines}.}
  \bibinfo{year}{2019}\natexlab{}.
\newblock \showarticletitle{Studying dawn-dusk asymmetries of {Mercury}'s
  magnetotail using {MHD-EPIC} simulations}.
\newblock \bibinfo{journal}{\emph{submitted to Journal of Geophysical Research:
  Space Physics}} (\bibinfo{year}{2019}).
\newblock


\bibitem[\protect\citeauthoryear{Daldorff, T{\'o}th, Gombosi, Lapenta, Amaya,
  Markidis, and Brackbill}{Daldorff et~al\mbox{.}}{2014}]%
        {daldorff2014two}
\bibfield{author}{\bibinfo{person}{Lars~KS Daldorff},
  \bibinfo{person}{G{\'a}bor T{\'o}th}, \bibinfo{person}{Tamas~I Gombosi},
  \bibinfo{person}{Giovanni Lapenta}, \bibinfo{person}{Jorge Amaya},
  \bibinfo{person}{Stefano Markidis}, {and} \bibinfo{person}{Jeremiah~U
  Brackbill}.} \bibinfo{year}{2014}\natexlab{}.
\newblock \showarticletitle{Two-way coupling of a global {Hall}
  magnetohydrodynamics model with a local implicit particle-in-cell model}.
\newblock \bibinfo{journal}{\emph{J. Comput. Phys.}}  \bibinfo{volume}{268}
  (\bibinfo{year}{2014}), \bibinfo{pages}{236--254}.
\newblock


\bibitem[\protect\citeauthoryear{Jordanova, Delzanno, Henderson, Godinez,
  Jeffery, Lawrence, Morley, Moulton, Vernon, Woodroffe,
  et~al\mbox{.}}{Jordanova et~al\mbox{.}}{2018}]%
        {jordanova2018specification}
\bibfield{author}{\bibinfo{person}{Vania~Koleva Jordanova},
  \bibinfo{person}{Gian~Luca Delzanno}, \bibinfo{person}{Michael~Gerard
  Henderson}, \bibinfo{person}{Humberto~C Godinez}, \bibinfo{person}{CA
  Jeffery}, \bibinfo{person}{Earl~Christopher Lawrence},
  \bibinfo{person}{Steven~Karl Morley}, \bibinfo{person}{John~David Moulton},
  \bibinfo{person}{Louis~James Vernon}, \bibinfo{person}{Jesse~Richard
  Woodroffe}, {et~al\mbox{.}}} \bibinfo{year}{2018}\natexlab{}.
\newblock \showarticletitle{Specification of the near-{Earth} space environment
  with {SHIELDS}}.
\newblock \bibinfo{journal}{\emph{Journal of Atmospheric and Solar-Terrestrial
  Physics}}  \bibinfo{volume}{177} (\bibinfo{year}{2018}),
  \bibinfo{pages}{148--159}.
\newblock


\bibitem[\protect\citeauthoryear{Lapenta, Markidis, Divin, Goldman, and
  Newman}{Lapenta et~al\mbox{.}}{2011}]%
        {lapenta2011bipolar}
\bibfield{author}{\bibinfo{person}{Giovanni Lapenta}, \bibinfo{person}{Stefano
  Markidis}, \bibinfo{person}{A Divin}, \bibinfo{person}{MV Goldman}, {and}
  \bibinfo{person}{DL Newman}.} \bibinfo{year}{2011}\natexlab{}.
\newblock \showarticletitle{Bipolar electric field signatures of reconnection
  separatrices for a hydrogen plasma at realistic guide fields}.
\newblock \bibinfo{journal}{\emph{Geophysical Research Letters}}
  \bibinfo{volume}{38}, \bibinfo{number}{17} (\bibinfo{year}{2011}).
\newblock


\bibitem[\protect\citeauthoryear{Lapenta, Pierrard, Keppens, Markidis,
  et~al\mbox{.}}{Lapenta et~al\mbox{.}}{2013}]%
        {lapenta2013swiff}
\bibfield{author}{\bibinfo{person}{Giovanni Lapenta}, \bibinfo{person}{Viviane
  Pierrard}, \bibinfo{person}{Rony Keppens}, \bibinfo{person}{Stefano
  Markidis}, {et~al\mbox{.}}} \bibinfo{year}{2013}\natexlab{}.
\newblock \showarticletitle{{SWIFF}: Space weather integrated forecasting
  framework}.
\newblock \bibinfo{journal}{\emph{Journal of Space Weather and Space Climate}}
  \bibinfo{volume}{3} (\bibinfo{year}{2013}), \bibinfo{pages}{A05}.
\newblock


\bibitem[\protect\citeauthoryear{Ma, Russell, T{\'o}th, Chen, Nagy, Harada,
  McFadden, Halekas, Lillis, Connerney, et~al\mbox{.}}{Ma
  et~al\mbox{.}}{2018}]%
        {ma2018reconnection}
\bibfield{author}{\bibinfo{person}{Yingjuan Ma}, \bibinfo{person}{Christopher~T
  Russell}, \bibinfo{person}{G{\'a}bor T{\'o}th}, \bibinfo{person}{Yuxi Chen},
  \bibinfo{person}{Andrew~F Nagy}, \bibinfo{person}{Yuki Harada},
  \bibinfo{person}{James McFadden}, \bibinfo{person}{Jasper~S Halekas},
  \bibinfo{person}{Rob Lillis}, \bibinfo{person}{John~EP Connerney},
  {et~al\mbox{.}}} \bibinfo{year}{2018}\natexlab{}.
\newblock \showarticletitle{Reconnection in the {Martian} Magnetotail:
  {Hall-MHD} With Embedded Particle-in-Cell Simulations}.
\newblock \bibinfo{journal}{\emph{Journal of Geophysical Research: Space
  Physics}} \bibinfo{volume}{123}, \bibinfo{number}{5} (\bibinfo{year}{2018}),
  \bibinfo{pages}{3742--3763}.
\newblock


\bibitem[\protect\citeauthoryear{Markidis, Henri, Lapenta, R{\"o}nnmark,
  Hamrin, Meliani, and Laure}{Markidis et~al\mbox{.}}{2014}]%
        {markidis2014fluid}
\bibfield{author}{\bibinfo{person}{Stefano Markidis}, \bibinfo{person}{Pierre
  Henri}, \bibinfo{person}{Giovanni Lapenta}, \bibinfo{person}{Kjell
  R{\"o}nnmark}, \bibinfo{person}{Maria Hamrin}, \bibinfo{person}{Zakaria
  Meliani}, {and} \bibinfo{person}{Erwin Laure}.}
  \bibinfo{year}{2014}\natexlab{}.
\newblock \showarticletitle{The fluid-kinetic particle-in-cell method for
  plasma simulations}.
\newblock \bibinfo{journal}{\emph{J. Comput. Phys.}}  \bibinfo{volume}{271}
  (\bibinfo{year}{2014}), \bibinfo{pages}{415--429}.
\newblock


\bibitem[\protect\citeauthoryear{Markidis, Lapenta, and Rizwan-uddin}{Markidis
  et~al\mbox{.}}{2010}]%
        {markidis2010multi}
\bibfield{author}{\bibinfo{person}{Stefano Markidis}, \bibinfo{person}{Giovanni
  Lapenta}, {and} \bibinfo{person}{Rizwan-uddin}.}
  \bibinfo{year}{2010}\natexlab{}.
\newblock \showarticletitle{Multi-scale simulations of plasma with {iPIC3D}}.
\newblock \bibinfo{journal}{\emph{Mathematics and Computers in Simulation}}
  \bibinfo{volume}{80}, \bibinfo{number}{7} (\bibinfo{year}{2010}),
  \bibinfo{pages}{1509--1519}.
\newblock


\bibitem[\protect\citeauthoryear{Markidis, Olshevsky, Sishtla, Chien, Laure,
  and Lapenta}{Markidis et~al\mbox{.}}{2018}]%
        {polypic}
\bibfield{author}{\bibinfo{person}{Stefano Markidis},
  \bibinfo{person}{Vyacheslav Olshevsky}, \bibinfo{person}{Chaitanya~Prasad
  Sishtla}, \bibinfo{person}{Steven W.~D. Chien}, \bibinfo{person}{Erwin
  Laure}, {and} \bibinfo{person}{Giovanni Lapenta}.}
  \bibinfo{year}{2018}\natexlab{}.
\newblock \showarticletitle{{PolyPIC}: The Polymorphic-Particle-in-Cell Method
  for Fluid-Kinetic Coupling}.
\newblock \bibinfo{journal}{\emph{Frontiers in Physics}}  \bibinfo{volume}{6}
  (\bibinfo{year}{2018}), \bibinfo{pages}{100}.
\newblock


\bibitem[\protect\citeauthoryear{Nishikawa}{Nishikawa}{1997}]%
        {nishikawa1997particle}
\bibfield{author}{\bibinfo{person}{K-I Nishikawa}.}
  \bibinfo{year}{1997}\natexlab{}.
\newblock \showarticletitle{Particle entry into the magnetosphere with a
  {Southward} interplanetary magnetic field studied by a three-dimensional
  electromagnetic particle code}.
\newblock \bibinfo{journal}{\emph{Journal of Geophysical Research: Space
  Physics}} \bibinfo{volume}{102}, \bibinfo{number}{A8} (\bibinfo{year}{1997}),
  \bibinfo{pages}{17631--17641}.
\newblock


\bibitem[\protect\citeauthoryear{Peng, Markidis, Laure, Johlander, Vaivads,
  Khotyaintsev, Henri, and Lapenta}{Peng et~al\mbox{.}}{2015a}]%
        {peng2015kinetic}
\bibfield{author}{\bibinfo{person}{Ivy~Bo Peng}, \bibinfo{person}{Stefano
  Markidis}, \bibinfo{person}{Erwin Laure}, \bibinfo{person}{Andreas
  Johlander}, \bibinfo{person}{Andris Vaivads}, \bibinfo{person}{Yuri
  Khotyaintsev}, \bibinfo{person}{Pierre Henri}, {and}
  \bibinfo{person}{Giovanni Lapenta}.} \bibinfo{year}{2015}\natexlab{a}.
\newblock \showarticletitle{Kinetic structures of quasi-perpendicular shocks in
  global particle-in-cell simulations}.
\newblock \bibinfo{journal}{\emph{Physics of Plasmas}} \bibinfo{volume}{22},
  \bibinfo{number}{9} (\bibinfo{year}{2015}), \bibinfo{pages}{092109}.
\newblock


\bibitem[\protect\citeauthoryear{Peng, Markidis, Vaivads, Vencels, Amaya,
  Divin, Laure, and Lapenta}{Peng et~al\mbox{.}}{2015b}]%
        {peng2015formation}
\bibfield{author}{\bibinfo{person}{Ivy~Bo Peng}, \bibinfo{person}{Stefano
  Markidis}, \bibinfo{person}{Andris Vaivads}, \bibinfo{person}{Juris Vencels},
  \bibinfo{person}{Jorge Amaya}, \bibinfo{person}{Andrey Divin},
  \bibinfo{person}{Erwin Laure}, {and} \bibinfo{person}{Giovanni Lapenta}.}
  \bibinfo{year}{2015}\natexlab{b}.
\newblock \showarticletitle{The formation of a magnetosphere with implicit
  particle-in-cell simulations}. In \bibinfo{booktitle}{\emph{15th ICCS)}}.
  \bibinfo{pages}{1178--1187}.
\newblock


\bibitem[\protect\citeauthoryear{Peng, Vencels, Lapenta, Divin, Vaivads, Laure,
  and Markidis}{Peng et~al\mbox{.}}{2015c}]%
        {peng2015energetic}
\bibfield{author}{\bibinfo{person}{Ivy~Bo Peng}, \bibinfo{person}{Juris
  Vencels}, \bibinfo{person}{Giovanni Lapenta}, \bibinfo{person}{Andrey Divin},
  \bibinfo{person}{Andris Vaivads}, \bibinfo{person}{Erwin Laure}, {and}
  \bibinfo{person}{Stefano Markidis}.} \bibinfo{year}{2015}\natexlab{c}.
\newblock \showarticletitle{Energetic particles in magnetotail reconnection}.
\newblock \bibinfo{journal}{\emph{Journal of Plasma Physics}}
  \bibinfo{volume}{81}, \bibinfo{number}{2} (\bibinfo{year}{2015}).
\newblock


\bibitem[\protect\citeauthoryear{Powell, Roe, Linde, Gombosi, and
  De~Zeeuw}{Powell et~al\mbox{.}}{1999}]%
        {powell1999solution}
\bibfield{author}{\bibinfo{person}{Kenneth~G Powell}, \bibinfo{person}{Philip~L
  Roe}, \bibinfo{person}{Timur~J Linde}, \bibinfo{person}{Tamas~I Gombosi},
  {and} \bibinfo{person}{Darren~L De~Zeeuw}.} \bibinfo{year}{1999}\natexlab{}.
\newblock \showarticletitle{A solution-adaptive upwind scheme for ideal
  magnetohydrodynamics}.
\newblock \bibinfo{journal}{\emph{J. Comput. Phys.}} \bibinfo{volume}{154},
  \bibinfo{number}{2} (\bibinfo{year}{1999}), \bibinfo{pages}{284--309}.
\newblock


\bibitem[\protect\citeauthoryear{T{\'o}th, Chen, Gombosi, Cassak, Markidis, and
  Peng}{T{\'o}th et~al\mbox{.}}{2017}]%
        {toth2017scaling}
\bibfield{author}{\bibinfo{person}{G{\'a}bor T{\'o}th}, \bibinfo{person}{Yuxi
  Chen}, \bibinfo{person}{Tamas~I Gombosi}, \bibinfo{person}{Paul Cassak},
  \bibinfo{person}{Stefano Markidis}, {and} \bibinfo{person}{Ivy~Bo Peng}.}
  \bibinfo{year}{2017}\natexlab{}.
\newblock \showarticletitle{Scaling the ion inertial length and its
  implications for modeling reconnection in global simulations}.
\newblock \bibinfo{journal}{\emph{Journal of Geophysical Research: Space
  Physics}} \bibinfo{volume}{122}, \bibinfo{number}{10} (\bibinfo{year}{2017}).
\newblock


\bibitem[\protect\citeauthoryear{T{\'o}th, Jia, Markidis, Peng, Chen,
  et~al\mbox{.}}{T{\'o}th et~al\mbox{.}}{2016}]%
        {toth2016extended}
\bibfield{author}{\bibinfo{person}{G{\'a}bor T{\'o}th},
  \bibinfo{person}{Xianzhe Jia}, \bibinfo{person}{Stefano Markidis},
  \bibinfo{person}{Ivy~Bo Peng}, \bibinfo{person}{Yuxi Chen}, {et~al\mbox{.}}}
  \bibinfo{year}{2016}\natexlab{}.
\newblock \showarticletitle{Extended magnetohydrodynamics with embedded
  particle-in-cell simulation of {Ganymede}'s magnetosphere}.
\newblock \bibinfo{journal}{\emph{Journal of Geophysical Research: Space
  Physics}} \bibinfo{volume}{121}, \bibinfo{number}{2} (\bibinfo{year}{2016}),
  \bibinfo{pages}{1273--1293}.
\newblock


\bibitem[\protect\citeauthoryear{Vencels, Delzanno, Johnson, Peng, Laure, and
  Markidis}{Vencels et~al\mbox{.}}{2015}]%
        {vencels2015spectral}
\bibfield{author}{\bibinfo{person}{Juris Vencels}, \bibinfo{person}{Gian~Luca
  Delzanno}, \bibinfo{person}{Alec Johnson}, \bibinfo{person}{Ivy~Bo Peng},
  \bibinfo{person}{Erwin Laure}, {and} \bibinfo{person}{Stefano Markidis}.}
  \bibinfo{year}{2015}\natexlab{}.
\newblock \showarticletitle{Spectral solver for multi-scale plasma physics
  simulations with dynamically adaptive number of moments}.
\newblock \bibinfo{journal}{\emph{Procedia Computer Science}}
  \bibinfo{volume}{51}, \bibinfo{number}{C} (\bibinfo{year}{2015}).
\newblock


\bibitem[\protect\citeauthoryear{Zhou, Toth, Jia, Chen, and Markidis}{Zhou
  et~al\mbox{.}}{2019}]%
        {newGanymede}
\bibfield{author}{\bibinfo{person}{Hongyang Zhou}, \bibinfo{person}{Gabor
  Toth}, \bibinfo{person}{Xianzhe Jia}, \bibinfo{person}{Yuxi Chen}, {and}
  \bibinfo{person}{Stefano Markidis}.} \bibinfo{year}{2019}\natexlab{}.
\newblock \showarticletitle{Embedded Kinetic Simulation of {Ganymede}'s
  Magnetosphere: Improvements and Inferences}.
\newblock \bibinfo{journal}{\emph{submitted to Journal of Geophysical Research:
  Space Physics}} (\bibinfo{year}{2019}).
\newblock


\end{thebibliography}

\end{document}